\documentclass[12pt]{article}
\usepackage[mathscr]{eucal}
\usepackage{amsmath,amsfonts,amssymb,amsthm}
\usepackage[matrix,arrow]{xy}
\usepackage[titletoc,title]{appendix}
\usepackage{cite}
\theoremstyle{definition}

\newcommand{\bref}[1]{\textbf{\ref{#1}}}

\def\be{\begin{equation}}
\def\ee{\end{equation}}
\def\ba{\begin{array}}
\def\ea{\end{array}}

\def\fnote#1#2{\begingroup\def\thefootnote{#1}\footnote{#2}\addtocounter
{footnote}{-1}\endgroup}

\numberwithin{equation}{section} \makeatletter
\@addtoreset{equation}{section}

\begin{document}

\begin{flushright}
FIAN-TD-2014-09 \\
\end{flushright}

\bigskip
\bigskip
\begin{center}

{\Large\textbf{Unitary Minimal Liouville Gravity\\ \vspace{4mm} and Frobenius Manifolds }}

\vspace{.8cm}

{\large{V. Belavin}$^{1,2}${\,}\fnote{*}{E-mail: belavin@lpi.ru}}

\bigskip
\bigskip
\bigskip

\begin{tabular}{ll}
$^{1}$~\parbox[t]{0.9\textwidth}{\normalsize\raggedright
{\it I.E. Tamm Department of Theoretical Physics, P.N. Lebedev Physical Institute, Leninsky ave. 53,
119991 Moscow, Russia}}\\
$^{2}$~\parbox[t]{0.9\textwidth}{\normalsize\raggedright
{\it Department of Quantum Physics, Institute for Information Transmission Problems, Bolshoy Karetny per. 19, 127994 Moscow, Russia}}
\end{tabular}\\

\bigskip
\bigskip
\bigskip

\begin{abstract}
We study unitary minimal models coupled
to Liouville gravity using Douglas string equation. Our approach is based
on the assumption that there exist an appropriate solution of the 
Douglas string equation and some special choice of the resonance transformation 
such that necessary selection rules of the minimal Liouville gravity are satisfied.
We use the connection with the Frobenius manifold structure.
We argue that the flat coordinates on the Frobenius manifold are the most
appropriate choice for calculating correlation functions. 
We find the appropriate solution of the Douglas string equation and show that it has
simple form in the flat coordinates. Important information
is encoded in the structure constants of the Frobenius algebra. We derive these structure constants
in the canonical coordinates and in the physically relevant domain in the flat coordinates.
We find the leading terms of the resonance transformation and express
the coefficients of the resonance transformation in terms
of Jacobi polynomials.
\end{abstract}

\end{center}

\newpage

\section{Introduction}
There are two independent approaches to 2D Liouville gravity \cite{Polyakov:1981rd}. The first
method is based on the continual integral over Riemann metrics. After gauge fixing,
it leads to the Liouville theory interacting with matter fields. If matter fields are
represented by some minimal CFT model, then
this theory is called minimal Liouville gravity (MLG).
The main object of study in MLG is correlation functions, which are given by
integrals over moduli of curves (with punctures) from differential
forms constructed through conformal blocks \cite{Belavin:1984vu}.
Another approach to 2D gravity is based on the discrete approximation of 2D surfaces
by graphs (studied in the double scaling limit). This approach is known as
matrix models (MM) of 2D gravity \cite{Kazakov:1985ea, Kazakov:1986hu, Kazakov:1989bc, Staudacher:1989fy, Brezin:1990rb, Douglas:1989ve, Gross:1989vs}.

One of the most important results concerning 2D gravity is the coincidence of
the spectra of gravitational dimensions in the MLG and MM approaches.
This observation led to the hypothesis  that these two approaches are
equivalent \cite{Gross:1989vs},\cite{Knizhnik:1988ak}.
But the connection 
is not obvious, because the results for correlation functions do not coincide.
As pointed out in \cite{Moore:1991ir},
the problem in establishing the correspondence between MM and MLG is due to the freedom
to add contact terms to the OPE. This effect leads
to mixing the coupling constants.
To identify the operators and the correlation functions, an explicit form
of the resonance relations between couplings is needed.
In \cite{Belavin:2008kv}, the relation between Liouville
gravity and MM was studied for the Lee--Yang series of minimal models (2,2p+1).
The explicit form of the resonance relations in terms of Legendre polynomials
was found. In \cite{Belavin:2013}, these results were
generalized to $(3,3p+p_0)$. It was shown that in this case, the basic requirements, i.e.,
absence of vacuum expectation values of physical operators, diagonal structure
of the two-point correlators, and satisfaction of the conformal fusion rules,
also leads to agreement between the two approaches on the level of the correlation functions.

The technique used in \cite{Belavin:2008kv},\cite{Belavin:2013}
was to some extent specific to the particular cases under consideration.
It was essentially based on the requirement that one of the parameters
is small ($q=2,3$). On the other hand,  in \cite{Belavin:2013},
the connection of (p,q) critical points of MLG with the Frobenius manifold structure was significantly clarified.
It was shown that the generating function of correlation numbers
represents the logarithm of the tau-function of the integrable hierarchy
related to some special Frobenious manifold. This connection can be used
to obtain an explicit representation of the free energy in terms of the tau-function.
The construction essentially requires knowing the structure constants of the Frobenius algebra.
With these achievements, it is  natural to tackle more general
examples of MLG.

In this work, we use the Douglas approach and the abovementioned connection
with the Frobenious manifold structure to study $(q+1,q)$ unitary models of MLG.
The rest of the paper is organized as follows. In Section \bref{sec:Douglas},
we briefly review the essence of the Douglas approach. We then introduce the Frobenious manifolds and
find the solution of the Douglas string equation relevant for MLG.
In this section, we also formulate the results for 
the structure constants of the Frobenius algebra.
In Section \bref{sec:corfun}, we study conformal selection rules and
the resonance relations. We apply the idea about the appropriate solution
of the string equation formulated in the preceding section for one- and two-point correlation functions
and discuss the resulting form of the resonance relations. Section \bref{sec:Concl} is for some concluding
remarks.

\section{Douglas approach and Frobenius manifolds}
\label{sec:Douglas}

The MM of 2D gravity  and the Douglas string equation approach \cite{Douglas:1989dd}
are widely studied in the literature (see, e.g., \cite{Moore:1991ir} and
the references therein). Below, we briefly review the ideas relevant to our discussion.

The MM exhibits an infinite set of multi-critical points. In the scaling limit
near the $(p,q)$ critical point, the singular part of the free energy $\cal F$
of the $p$-matrix model is constructed in terms of the solution of the string equation
\be
[\hat{P},\hat{Q}]=1,
\label{streq}
\ee
where $\hat{P}$ and $\hat{Q}$ are two differential operators,
\begin{align} \label{Q def}
\hat{Q} &= \big(\frac{d}{d x}\big)^q+\sum_{\alpha =1}^{q-1}u_{\alpha}(x)\big(\frac{d}{d x}\big)^{q-\alpha -1} ,\\
\label{P def}
\hat{P} &= \frac{p+q}{ q^2} \hat{Q}^{\frac{p}{q}}_{\,\,+} + \sum_{\alpha=1}^{q-1}\sum_{k=1}  \frac{k q+ \alpha}{ q^2} t_{k,\alpha} \hat{Q}^{k+ \frac{\alpha}{q}-1}_{\,\,+},
\end{align}
and the plus sign means that only nonnegative powers of $d/dx$ are taken
in the series expansion of the pseudo-differential operator.
The constants $t_{k,\alpha}$ are usually called ``times". This choice of the
basis in the parameter space is called the KdV frame.

The solution $u^*_{\alpha}(x)$ of the string equation \eqref{streq}
is related to the nonsingular part of the free energy
\be  \label{douglas free energy}
\frac{\partial^2 {\cal F}}{\partial x^2} = u_1^*(x).
\ee
The main conjecture about the connection between the Matrix Models of 2D gravity and
the continuous Liouville approach is that for the appropriate solution
of the string equation the singular part of the free energy $\cal F$
coincides with the generating function of correlators of different topologies
in the $(p,q)$ MLG.\footnote{The generating function can be thought of as
a partition function of the perturbed MM coupled to Liouville gravity,
which is why the generating function is called sometimes the partition function.}
One further point is worthy of note. It may seem from the above
simple definition that this construction comprises exhaustive information
about the theory and that the remaining questions are quite technical and
straightforward. But the problem is that if the partition function is the
subject of the Douglas string equation, then there still remains much ambiguity
related to the choice of the point in the space of theories (the point in the
parameter space) and the choice of the appropriate coordinates in this space
such that this particular setup leads to correct answers for the correlation
functions. The transformation from the initial coordinates $t_{k,\alpha}$ to
the appropriate ``Liouville'' coordinates $\lambda_{mn}$ is known as a
resonance transformation \cite{Moore:1991ir} (it is discussed in detail below).
Hence, there is a conceptual question of clarifying  the meaning of these choices
as well as the meaning of the other possible solutions of \eqref{streq} and
the relations between them.

In what follows we restrict our attention to the planar limit corresponding to the spherical topology.
In this case, we are dealing with the dispersionless limit of the string equation
when $d/dx$ is replaced with the variable $y$ and the commutator is replaced with the Poisson bracket.
In particular, the operator $\hat{Q}$ transforms to the polynomial
\be\label{Q}
Q(y)=y^q+u_1 y^{q-2}+u_2 y^{q-3}+\cdots+u_{q-1}.
\ee
In the planar limit, we are left with the system of first-order PDEs and
the Douglas string equation takes the simple form \cite{Belavin:2013},\cite{Ginsparg:1990zc}
\be \label{action principle}
\frac{\partial S}{ \partial u_k} = 0,
\ee
where the so-called action is
\be \label{action ambigious}
S[u_{\alpha}(x)]  =-\underset{y=\infty}{\text{res}}
\bigg (Q^{\frac{p+q}{q}}+\!\!\!\!\sum_{1\leq n\leq m \leq q-1}\!\!\!\! t_{mn} (\mu,\lambda)\,\,Q^{\frac{p m-q n}{q}} \bigg).
\ee
The singular part of the partition function is expressed \cite{Belavin:2013} in terms of
\eqref{action ambigious} as an integrated (closed) one-form
\begin{eqnarray}
\mathcal{F}=\frac 12 \int_0^{{\bf u}^*} \widetilde{C}_{k}^{i j}
\frac{\partial S}{\partial u^{i}}\frac{\partial S}{\partial u^{j}} d u^{k},
\label{frenerg}
\end{eqnarray}
where ${\bf u}^*$ is a suitable solution of the string equation \eqref{action principle}
and $\widetilde{C}_{k}^{i j}$ are the structure
constants\footnote{We reserve C without the tilde for the flat coordinate system.} of the Frobenius algebra
discussed in the next section.   Hence, according to our basic idea,
this function gives the generating function of $(p,q)$ MLG
\be
\mathcal{F}= \langle \exp \sum_{m,n} \lambda_{m,n} O_{m,n} \rangle,
\ee
where the Liouville coupling constants $\lambda_{m,n}$ are expressed in terms
of the times $t_{mn}$ via the resonance transformation. Using the Douglas approach
to solve MLG thus requires three main ingredients: the appropriate solution of the
string equation, the resonance relations between times in the KdV frame and
Liouville couplings, and, finally, the structure constants of the Frobenius algebra.

\subsection{Frobenius algebra structure constants}
\label{sec:strconst}
In general, a Frobenius algebra $A$ is a commutative algebra with unity equipped
with a nondegenerate invariant pairing $A\otimes A \rightarrow \mathbb{C}$.
The invariance means that for any three vectors $a,b,c$ in $A$:
\begin{equation}\label{inva}
(a\cdot b, c)=(a, b\cdot c).
\end{equation}

The Frobenius manifold $M$ \cite{Dubrovin:1992dz} is defined as a Riemannian
manifold  with the flat metric $g_{\bf v}$  compatible with the additional
structure of the Frobenius algebra $A$ associated with each point  ${\bf v}$
(or with each tangent space $T_{\bf v} M$).
Let $v^1,\dots, v^n$ be the flat coordinates on an $n$-dimensional Frobenius
manifold $M$. The algebra $A$ is identified with the tangent space $T_{\bf v}M$ by
\be
e_{\alpha} \leftrightarrow \frac{\partial}{\partial v^{\alpha}},
\ee
where $\alpha$ runs from $1$ to $n$. We hence have the following multiplication
rule for the tangent vectors at any point ${\bf v}\in M$:
\be
\frac{\partial}{\partial v^{\alpha}}\cdot \frac{\partial}{\partial v^{\beta}}=
c_{\alpha\beta}^{\gamma}({\bf v}) \frac{\partial}{\partial v^{\gamma}}.
\ee
The axiom
\begin{equation}\label{fro4}
\frac{\partial}{\partial v^{\delta}} c_{\alpha\beta\gamma}({\bf v})=
\frac{\partial}{\partial v^{\gamma}} c_{\alpha\beta\delta}({\bf v})
\end{equation}
defines the structure of the Frobenius manifold (see, e.g., \cite{Belavin:2013}
for more details about properties of Frobenius manifolds).

In Liouville gravity, we deal with a particular example of the Frobenius manifold
where the corresponding algebra
is defined as
an algebra of polynomials $C[y]$ modulo the polynomial $Q'=dQ/dy$, where $Q$ is
defined in \eqref{Q}. The parameters $u_i$ in \eqref{Q} are the coordinates on
the Frobenius manifold. We call these coordinated canonical coordinates.
In the basis $e_1,\dots,e_{q-1}$, where $e_{k}=y^{q-1-k}$, the algebra looks like
\begin{eqnarray}\label{algebra}
e_{i}e_{j}=\widetilde{C}_{ij}^{k}e_{k} \!\!\!\mod Q'\,,
\end{eqnarray}
and the metric $g_{ij}$ is given by
\be \label{metricu}
g_{ij} :=\underset{y=\infty}{\text{res}}
\frac{e_{i}e_{j}}{ Q'}.
\ee
The structure constants in \eqref{algebra} are exactly those that appeared in \eqref{frenerg}.
The metric $g_{ij}$ (and its inverse) is used to raise and lower the indices. Using this,
we find
\be \label{strconstu1}
 \widetilde{C}_{ijk}=\underset{y=\infty}{\text{res}} \frac{e_{i}e_{j}e_{k}}{ Q'}.
\ee
In particular, the metric itself is expressed in terms of the structure constants
$g_{ij}=\widetilde{C}_{i\,j\, q-1}$. In each particular case, i.e., for a given pair $(p,q)$,
the structure constants can be directly evaluated from the definition \eqref{strconstu1}.
For example, for $q=5$, we find
{\small
$$
\begin{tabular}{|l|l|l|l|}
$\widetilde{C}_1^{ij}$&$\widetilde{C}_2^{ij}$&$\widetilde{C}_3^{ij}$&$\widetilde{C}_4^{ij}$\\
\hline
\begin{tabular}{cccc}\small
 5 & 0 & 0 & 0 \\
 0 & -$3 u_1$ & -$2 u_2$ & -$u_3$ \\
 0 & -$2 u_2$ & -$u_3$ & 0 \\
 0 & -$u_3$ & 0 & 0\\
\end{tabular}
&
\begin{tabular}{cccc}\small
 0 & 5 & 0 & 0 \\
 5 & 0 & 0 & 0 \\
 0 & 0 & -$2 u_2$ & -$u_3$ \\
 0 & 0 & -$u_3$ & 0\\
\end{tabular}
&
\begin{tabular}{cccc}\small
 0 & 0 & 5 & 0 \\
 0 & 5 & 0 & 0 \\
 5 & 0 &  $3 u_1$ & 0 \\
 0 & 0 & 0 & -$u_3$\\
\end{tabular}
&
\begin{tabular}{cccc}\small
 0 & 0 & 0 & 5 \\
 0 & 0 & 5 & 0 \\
 0 & 5 & 0 &  $3 u_1$ \\
 5 & 0 &  $3 u_1$ &  $2 u_2$\\
\end{tabular}
\end{tabular}\;.
$$}

\noindent
It is worth noting that such a simple linear dependence characterizes the
structure constants in $u$ coordinates only if two indices are raised.
For all indices down, the dependence is nonlinear. This example demonstrates
the following properties of the structure constants $\widetilde{C}_{ij}^{k}$.
For a given index $i$, it has a block diagonal structure, and regarded as a
function of $j$ and $k$, it depends on the combination $j+k$, which means
that its counter-diagonal elements are the same.
These properties together with the explicit calculations for particular
values of the parameter $q$ (one of which is shown above) suggest the
general form of the structure constant in ${\bf u}$ coordinates:
\begin{eqnarray}
\label{strconst}
\widetilde{C}_{k}^{ij}=(q+k-i-j+1) u_{i+j-k-2} \widetilde{\Theta}(i,j,k),
\end{eqnarray}
where $\widetilde{\Theta}(i,j,k)$ is given by
\begin{eqnarray}
\widetilde{\Theta}(i,j,k)=
\begin{cases}\phantom{-}1 & \text{if}\quad i,j\leq k \quad \text{and} \quad i+j > k,
\\-1 & \text{if}\quad i,j> k \quad \text{and} \quad i+j \leq k+q, \\
\phantom{-}0 & \text{otherwise.} \end{cases}
\end{eqnarray}
In \eqref{strconst}, we imply that $u_{-1}=1,u_0=0$. So far, we have not
found a proof of the result \eqref{strconst}. It therefore remains
a conjecture verified up to $q=15$.

The result \eqref{strconst} allows writing the partition function \eqref{frenerg}
in the form
\begin{eqnarray}\label{F}
\mathcal{F}=\frac 12 \int_0^{{\bf u}^*} \widetilde{\Omega}_i d u^{i},
\end{eqnarray}
where
\begin{eqnarray}
 \widetilde{\Omega}_i = q \!\!\!\sum_{k,j}^{k+j=i+1}
 \frac{\partial S}{\partial u_{k}}\frac{\partial S}{\partial u_{j}}+
\bigg[\,\sum_{l\geq i+3}^{2i}\,\sum_{k,j\leq i}^{k+j=l}
-\sum_{l\geq 2i+2}^{i+q}\,\sum_{k,j> i}^{k+j=l}\, \bigg](q+i-l+1)u_{l-i-2}
\frac{\partial S}{\partial u_{k}}\frac{\partial S}{\partial u_{j}}.
\end{eqnarray}
Hence, the found structure constants \eqref{strconst} together with
the explicit form of the resonance relations, if it were known, could lead to
a rather simple expression for the generating function of correlators.
But one piece of data is missing: Which solution
of the string equation is related to minimal models coupled to gravity?
We discuss this question in the next section.

\subsection{Flat coordinates}

It was argued in \cite{DiFrancesco1990} that only one of the solutions
of the string equation is physical, i.e., it leads to zero one-point correlators.
Hence, we just have to find the correct one. In the $q=2$ case, there is only
one solution (modulo some physically irrelevant phase factors). For $q=3$,
when there are two coordinates $u_1$ and $u_2$, the situation is more subtle
because the string equations have now a number of essentially different
solutions. It was found in \cite{Belavin:2013} that there exists only one
particular solution with the necessary properties.
The correctness of the choice was verified by computing the correlators.
As a bonus, it turned out that the integration contour can be taken along
one of two axes because $u^*_2=0$ (for all couplings equal to zero) for this particular solution. We recall
that the integral is independent of the contour because the integrand is a closed one-form.

For $q>3$, the situation becomes more complicated.
The computations show that unlike the $q=2,3$ cases,
the roots of the string equation with only one nonzero component\footnote{Here the couplings equal to zero.} are nonphysical
in the sense that using these roots in the expression \eqref{F} for the partition function,
we cannot satisfy one of the basic requirements that one-point functions of physical fields
are equal to zero. The consideration in the canonical coordinates now becomes more involved
because the contour of the integration does not go along one of the axes. 
In brief, we conclude
that the appropriate physical solution of the Douglas string equation corresponding
to the general $(p,q)$ MLG for an arbitrary  value of the parameter $q$ seems complicated
in the canonical coordinates $u_i$.

Our idea is to find a transformation from the canonical coordinates to
some new coordinates such that  the physical solution would have the
simplest form in the new coordinate system. We find that there
exist coordinates $v_i$ such that the physical solution in the zeroth order
of the expansion in the coupling constants becomes $(v_1,0,0,\dots)$, i.e.,
$v_{k>1}=0$, similarly to  what happens in the canonical coordinates for $q=2,3$.
Our calculations for different values of the parameter $q$ led to the
conclusion that the correct choice of the coordinates coincides
with the choice of the flat coordinates on the Frobenius manifold.
The explicit form of the transformation from the canonical to the flat coordinates is
given by the equality
\begin{equation}\label{transform}
y=Q^{\frac 1q}(y)-
\frac{1}{q}\bigg(\frac{v_1}{Q^{\frac 1q}(y)}+\frac{v_2}{Q^{\frac 2q}(y)}+\dots+\frac{v_{q-1}}{Q^{\frac{q-1}q}(y)}\bigg)
+\mathcal{O}\bigg(\frac{1}{Q^{\frac{q+1}q}(y)}\bigg),
\end{equation}
where $Q(y)$ is defined in \eqref{Q}.

The flat coordinates have a number of useful properties.
First, the metric in these coordinates is constant. It has the simple form
\begin{equation}
\eta_{\alpha\beta}=\delta_{\alpha+\beta,q}.
\end{equation}
The flat coordinates are expressed explicitly in terms of $u_i$:
\begin{equation}
v_i=\theta_{i,0}.
\label{v_i}
\end{equation}
Here, $\theta_{i,k}$ ($k\in \mathbb{N}$) is an important object of the
Frobenius manifold structure which is relevant for the existence of
infinite number of commuting flows and for the relation with
the integrable hierarchies. (It represents a one-parameter
family of deformed flat connections on the Frobenius manifold; see,
e.g., \cite{Belavin:2013} for more details). It is given by
\begin{equation}
\theta_{\alpha,k}=-c_{\alpha,k} \underset{y=\infty}{\text{res}} Q^{k+\frac{\alpha}{q}}(y)
\label{theta}
\end{equation}
with
\begin{equation}
c_{\alpha,k}^{-1}=\bigg(\frac{\alpha}{q}\bigg)_{k+1},
\end{equation}
where $(a)_n=\Gamma(a+n)/\Gamma(a)$ is the Pochhammer symbol.
The explicit form of the polynomial $Q$, evaluated in the flat coordinates on the
solution of the string equation for the couplings equal to zero, is
\begin{equation}
Q(y)|_{\vec{v}_0}=y^q+q\sum_{k}^{2k-1<q}
\left( {\begin{array}{*{20}c} q-k-1 \\ k-1 \\ \end{array}} \right)\bigg(\frac{v_1}{q}\bigg)^k y^{q-2k}.
\end{equation}
This expression is exactly the one that appeared in the planar limit
of the $Q$ operator considered in \cite{DiFrancesco1990}.
The connection with the flat coordinates clarifies the geometric
meaning of this choice of the $Q$ operator.

Hence, we find that the string equation has one particular solution,
denoted by ${\bf{v}}^*$ in what follows, that is equal to
$\{v_1^*,0,0,\dots\}$ in the zeroth order in coupling constants.
It is shown in the next section that for this solution, the one-point
correlation functions are zero and the two-point functions satisfy
the diagonality requirement.

The structure constants in the flat coordinates are given by
\begin{equation}
C_{\alpha\beta\gamma}=-q\underset{y=\infty}{\text{res}}\frac{\frac{\partial Q(y)}{\partial v^\alpha}
\frac{\partial Q(y)}{\partial v^\beta}\frac{\partial Q(y)}{\partial v^\gamma}}{Q'(y)}.
\end{equation}
In particular, because $C_{1\alpha\beta}=\eta_{\alpha\beta}$, we obtain
\begin{equation}
C_{1}^{\alpha\beta}=\delta_{\alpha+\beta,q} \qquad \text{and}
\qquad C_{\alpha}^{q-1,\beta}=\delta_{\alpha,\beta}.
\label{C_properties}
\end{equation}
Finally, we note that because of the properties of the flat metric
$\eta_{\alpha\beta}$, we obtain the useful relation
\begin{equation}
v^{\alpha}=v_{q-\alpha}.
\end{equation}

Although we have obtained explicit structure constants in the canonical coordinates,
we have not yet found the general answer for the structure constants
in the flat coordinates. Nevertheless, taking the form of the
generating function into account, we can note that to analyze one- and
two-point correlation functions, we must know the structure constants
in the flat coordinates only on the line $v_{k>1}=0$. This becomes clear
when we take into account that the contour independence of the
expression \eqref{Z} can be used to choose it along $v_1$ and also that for
one- and two-point correlation functions, the derivatives act only on the
integrand.

To find the structure constant, we  perform the coordinate transformation
with the expression \eqref{strconst}\footnote{We use Greek and Latin
alphabets for the respective $\bf{v}$ and $\bf{u}$ coordinate systems.}
\be
C_{\alpha\beta\gamma} =\frac{\partial u^i}{\partial v^{\alpha}}
\frac{\partial u^j}{\partial v^{\beta}}\frac{\partial u^k}{\partial v^{\gamma}} \widetilde{C}_{ijk}.
\ee
This allows obtaining an explicit form of the expansion in the vicinity
of $\bf{v}^*$. We note that in the $n$-point correlation function, the contribution
gives a maximum $(n-3)$ terms in this expansion. The explicit results for
the first terms is presented below.
To represent the answer compactly, we introduce the function $\Theta_{A,B}(x)$
such that $\Theta_{A,B}(x)=1$ if $x\in [A,B]$ and is zero otherwise.
The result for the structure constant in the flat coordinates
on the line $v_{i>0}=0$ is
\begin{eqnarray}\label{strconstflat}
&C_{\alpha\beta\gamma}= \Theta_{1,q-1}(\alpha+\beta-\gamma)
\big(\!\!-\frac{v_1}{q}\big)^{\frac{\alpha+\beta+\gamma-q-1}{2}} \text{ if} \quad
\frac{\alpha+\beta+\gamma-q-1}{2}\in\mathbb{N},\text{ otherwise 0}.
\end{eqnarray}
In \eqref{strconstflat}, we assume the ordering $\alpha\geq\beta\geq\gamma$.
Because this tensor is symmetric, this information gives the complete answer.
We give some details of the derivation in Appendix~A.

\section{Correlation functions}
\label{sec:corfun}

Our aim in this section is to determine the correlation functions in MLG
using the special properties of the flat coordinates. All required information
concerning these correlators is encoded in the partition function.
In the flat variables,
\begin{equation}
\mathcal{F}=\frac 12 \int_0^{\bf{v}^*} C_{\alpha}^{\beta\gamma} \frac{\partial S}{\partial v^{\beta}}
 \frac{\partial S}{\partial v^{\gamma}} d v^{\alpha},
\label{Z}
\end{equation}
where $S$ defined in \eqref{action ambigious} is now written in the flat
coordinates. We use the resonance relations to satisfy the MLG selection
rules. Explicitly, the resonance transformation is
\begin{align}
t_{mn}\,=\,\,&\lambda_{mn}+\!\!\!\!\!\sum_{m_1,n_1}^{\delta_{m_1n_1}\leq
\delta_{mn}} \!\!\!\!A^{m_1n_1}_{mn}\mu^{\delta_{mn}-\delta_{m_1n_1}} \lambda_{m_1n_1}
+\nonumber\\
&+\!\!\!\!\sum_{m_1,n_1,m_2,n_2}^{\delta_{m_1n_1}+\delta_{m_2n_2}\leq
\delta_{mn}} \!\!\!\!\!\!A^{m_1n_1,m_2n_2}_{mn}\mu^{\delta_{mn}-
\delta_{m_1n_1}-\delta_{m_2n_2}} \lambda_{m_1n_1}\lambda_{m_2n_2}+\dots,
\label{coupling}
\end{align}
where $\lambda_{mn}$ are the Liouville coupling constants and the
constants $A^{m_1n_1}_{mn},A^{m_1n_1,m_2n_2}_{mn},\dots$ are to be defined
from the conformal and fusion selection rules. We note that the specific
property of the unitary series is the absence of resonance terms of the form $\mu^N$.
The gravitational dimensions $\delta_{mn}$ for the unitary MM
$M(q+1,q)$ are given by
\begin{equation}
\delta_{mn}=\frac{2q+1-|(q+1)m-q n|}{2q}.
\end{equation}
In particular, $\lambda_{11}=\mu$ is the cosmological constant, and
$\lambda_{mn}\sim \mu^{\delta_{mn}}$. In what follows, $\lambda_{mn}$ is used
only for $(m,n)\neq(1,1)$.
By regrouping the terms, we can write the expansion of the action
\begin{equation}\label{actionlambda}
S=S^{(0)}+\sum_{m,n} \lambda_{mn}\, S^{(mn)}+
\sum_{m_1,n_1,m_2,n_2} \lambda_{m_1n_1}\lambda_{m_2n_2}\, S^{(m_1n_1,m_2n_2)}+\ldots.
\end{equation}
The correlation numbers in the Liouville frame are related to the
coefficients in the coupling constant decomposition of the partition function
\begin{equation}
{\cal F}[t(\lambda)]=Z_0+\sum_{m_1,n_1} \lambda_{m_1n_1} Z_{m_1n_1}+
\sum_{m_1,n_1}\sum_{m_2,n_2} \lambda_{m_1n_1}\lambda_{m_2n_2} Z_{m_1n_1,m_2n_2}+\ldots.
\end{equation}
In other words, the correlation numbers are expressed in the form
\be \label{corrnumbers}
\langle O_{m_1n_1}\dots O_{m_Nn_N}\rangle =  Z_{m_1n_1,..,m_Nn_N}=
\frac{\partial}{ \partial \lambda_{m_1n_1}} \dots
\frac{\partial}{ \partial \lambda_{m_Nn_N} } \Bigg|_{\lambda =0 } {\cal F}[t(\lambda)].
\ee

One comment concerning dimensional analysis is in order. It is convenient to
use their dimensionless counterparts instead of considering the dimensional quantities.
This is achieved by changing variables $v_1\rightarrow v_1/v_{1}^*$. In what follows,
we let the same letter denote the new dimensionless variable.
Omitted dimensional factors have the form of some powers of $v_{*1}$.
They can be easily reconstructed using the homogeneity of the polynomial $Q$.
These factors, of course, do not contribute to the real
physical quantities, which are independent of the normalization.

\subsection{One-point functions}

In what follows, our basic tool is the recurrence relation \cite{Belavin:2013}
\begin{equation}
\frac{\partial^2 \theta_{\lambda,k+1}}{\partial v_{\alpha}\partial v_{\beta}}= C_{\gamma}^{\alpha\beta}
\frac{\partial \theta_{\lambda,k}}{\partial v_{\gamma}}.
\label{Recursion}
\end{equation}
The consideration is based on the following statement.
On the line $v_{i>1}=0$,
\begin{eqnarray}
\begin{cases}
k\text{ even}:&
\frac{\partial \theta_{\lambda,k}}{\partial v_{\alpha}}=\delta_{\lambda,\alpha} \,
x_{\lambda,k}\,\big(-\frac{v_1}{q}\big)^{\frac{k}{2}q},
\\
k\text{ odd}:&
\frac{\partial \theta_{\lambda,k}}{\partial v_{\alpha}}=\delta_{\lambda,q-\alpha} \,
y_{\lambda,k}\,\big(-\frac{v_1}{q}\big)^{\frac{k-1}{2}q+\lambda},
\end{cases}
\label{Lemma3}
\end{eqnarray}
where
\begin{eqnarray}\label{xy}
x_{\lambda,k}=\frac{1}{\big(\frac{\lambda}{q}\big)_{\frac{k}{2}}
\big(\frac{k}{2}\big)!}
\qquad\text{and}\qquad
y_{\lambda,k}=-\frac{1}{\big(\frac{\lambda}{q}\big)_{\frac{k+1}{2}}
\big(\frac{k-1}{2}\big)!}.
\label{Lemma31}
\end{eqnarray}
The proof is given in Appendix B.

The first requirement to be satisfied is the zero expectation values of
physical operators (except the unity operator). For the ${\bf v}^*$ solution,
we can take the contour of the integration along the $v_1$ axis.
Keeping in mind the comment at the end of the preceding section, we write
the one-point  function in the form
\begin{eqnarray}
\langle O_{mn} \rangle=\int_0^{1} C_{q-1}^{\beta\gamma} \frac{\partial S^{(0)}}{\partial v^{\beta}}
\frac{\partial S^{(mn)}}{\partial v^{\gamma}} d v_{1}.
\label{Omn}
\end{eqnarray}
Because of the string equation, the derivative in \eqref{Omn} acts only
on the integrand leading to $S^{(mn)}$, which is the first term  of the
series expansion of the action \eqref{actionlambda} in the coupling constants
\begin{equation}\label{Smn0}
S^{(mn)}=\underset{y=\infty}{\text{res}}Q^{\frac{(q+1)m-q n}{q}}+\!\!\!\!\sum_{m_1,n_1}^{\delta_{m_1n_1}<\delta_{mn}}
A_{m_1n_1}^{mn}\,\, \mu^{\delta_{mn}-\delta_{m_1n_1}} \,\,
\underset{y=\infty}{\text{res}}Q^{\frac{(q+1)m_1-q n_1}{q}},
\end{equation}
where $A_{m_1n_1}^{mn}$ are the coefficients in the resonance relations of
the coupling constants. Using \eqref{theta}, we find
\begin{equation}\label{S0}
S^{(0)}=\underset{y=\infty}{\text{res}}\bigg[Q^{\frac{2q+1}{q}}+\mu Q^{\frac{1}{q}}\bigg]=
-\frac{ \theta_{1,2}}{c_{1,2}}-\mu\frac{\theta_{1,0}}{c_{1,0}}.
\end{equation}
Taking \eqref{Lemma3} into account, we find that
$\frac{\partial S^{(0)}}{\partial v^{\beta}}=0$ for $\beta\neq q-1$.  Hence,
\begin{eqnarray}
\langle O_{mn}\rangle
=\int_0^{1} C_{q-1}^{q-1,\gamma} \frac{\partial S^{(0)}}{\partial v^{q-1}}
\frac{\partial S^{(mn)}}{\partial v^{\gamma}} d v_{1}=
\int_0^{1}\frac{\partial S^{(0)}}{\partial v_{1}}
\frac{\partial S^{(mn)}}{\partial v_{1}} d v_{1},
\label{Zmn1}
\end{eqnarray}
where we use $C_{\alpha}^{q-1,\beta}=\delta_{\alpha\beta}$ and lower the indices.

Now, the question is for which pairs $(m,n)$ the factors
$\frac{\partial S^{(mn)}}{\partial v_{1}}$ are nonzero. Using \eqref{Lemma3}
and \eqref{Smn0}, we find that for $m-n=2k+1$ with integer $k$,
the corresponding gravitational dimension of the one-point function
 $\langle O_{mn} \rangle$ is
\begin{equation}
[\langle O_{mn} \rangle]=\frac{2q+1}{q}-\delta_{mn}=(k+1),
\end{equation}
i.e., is analytic in $\mu$ and should not be considered \cite{Belavin:2013}. The second option is
$m-n$ even and $m=1$. Because $n\leq m$,  the only possible pair is $(m,n)$,
which corresponds to the unity operator.

It is instructive to obtain an explicit answer for the partition function
\begin{equation}
Z_{0}=\frac{1}{2}\int_0^{1} d v_1
\bigg(\frac{\partial S^{(0)}}{\partial v_1}\bigg)^2,
\label{Z0}
\end{equation}
where
\begin{equation}
\frac{\partial S^{(0)}}{\partial v_1}=
-\frac{1}{c_{1,2}} \frac{\partial \theta_{1,2}}{\partial v_1}-\frac{\mu}{c_{1,0}}.
\end{equation}
The partial derivative $\frac{\partial \theta_{1,2}}{\partial v_1}$
can be easily found from \eqref{Recursion}:
\begin{equation}
\frac{\partial \theta_{1,2}}{\partial v_1}=\bigg(-\frac{1}{q}\bigg)^{q-2}\frac{v_1^q}{q}.
\end{equation}
From the string equation $\frac{\partial S^{(0)}}{\partial v_1}\bigg{|}_{v_1=v_{1}^*}\!\!\!\!=0$,
we find that the cosmological constant, measured in $(v_{1}^*)^q$, is
\begin{equation}
\mu =- q\bigg(\!\!\!-\frac{1}{q}\bigg)^q  \frac{c_{1,0}}{c_{1,2}}.
\end{equation}
The integral \eqref{Z0} can be evaluated explicitly:
\begin{equation}
Z_{0}=\frac{(-\frac{1}{q})^{2q-4}}{ c_{1,2}^{2}}  \frac{1}{(1+q)(1+2q)}.
\label{Z0result}
\end{equation}

We conclude that nonzero one-point functions appear only for the unity operator and
for other operators having nonsingular gravitational dimensions. All singular
one-point functions are automatically  zero. We also note that when the flat
variables are used on the level of one-point functions, the first order coefficients in the
resonance expansion remain undetermined because the requirement of the absence
of one-point expectation values is satisfied automatically. This is a specific
property of the flat coordinate system. Hence, the explicit form of the resonance
relation should be fixed from the restrictions arising on the levels of
higher-point correlators.

\subsection{Two-point functions}
At the two-point level, the derivatives again act only  on the integrand.
Therefore, the result is
\begin{equation}
\langle O_{m_1n_1}O_{m_2n_2}\rangle= \int_0^{1} d v_1 C_{q-1}^{\alpha\beta}
\frac{\partial S^{(m_1n_1)}}{\partial v^{\alpha}} \frac{\partial S^{(m_2n_2)}}{\partial v^{\beta}}+
\int_0^{1} d v_1 C_{q-1}^{\alpha\beta}\frac{\partial S^{(0)}}{\partial v^{\alpha}}
\frac{\partial S^{(m_1n_1,m_2n_2)}}{\partial v^{\beta}}.
\end{equation}
It follows from the same arguments as in the one-point case that the second
integral vanishes. Hence,
\begin{equation}
\langle O_{m_1n_1}O_{m_2n_2}\rangle= \int_0^{1} d v_1 C_{q-1}^{\alpha\beta}
\frac{\partial S^{(m_1n_1)}}{\partial v^{\alpha}} \frac{\partial S^{(m_2n_2)}}{\partial v^{\beta}},
\end{equation}
which because of \eqref{strconstflat} can be written as
\begin{equation}\label{Z12}
\langle O_{m_1n_1}O_{m_2n_2}\rangle= \sum_{\gamma=1}^{q-1} (-q)^{1-\gamma}
\int_0^{1} d v_1\, v_1^{\gamma-1} \,
\frac{\partial S^{(m_1n_1)}}{\partial v_{\gamma}}\, \frac{\partial S^{(m_2n_2)}}{\partial v_{\gamma}}.
\end{equation}
Our next step is to derive the general form of the partial derivatives
$\frac{\partial S^{(mn)}}{\partial v_{\gamma}}$.
We find the following consequence of \eqref{Lemma3} for
$\frac{\partial S^{(mn)}}{\partial v_{\alpha}}$:
\begin{eqnarray}
\begin{cases}
(m-n)\text{ even}:&
\frac{\partial S^{(mn)}}{\partial v_{\alpha}}\sim
 \,\delta_{m,\alpha}
\bigg(v_1^{\frac{m-n}{2}q}\!+\!\dots\bigg),
\\
(m-n)\text{ odd}:&
\frac{\partial S^{(mn)}}{\partial v_{\alpha}}\sim \delta_{m,q-\alpha}\bigg( v_1^{\frac{m-n-1}{2}q+m}+\dots\bigg),
\end{cases}
\label{Smn_alpha0}
\end{eqnarray}
where dots denote terms involving the cosmological constant $\mu$.
We now consider the two point functions \eqref{Z12}.
It follows from the dimensional analysis that
\begin{equation}
[\langle O_{m_1n_1}O_{m_2n_2} \rangle]=\frac{2q+1}{q}\!-\!\delta_{m_1n_1}\!-\!\delta_{m_2n_2}
=\frac{m_1-n_1}{2}+\frac{m_2-n_2}{2}+\frac{m_1+m_2}{2q}.
\end{equation}
According to \eqref{Z12} and \eqref{Smn_alpha0},  there are three different cases.
In the case where one of the fields is even while another is odd
(for example, $m_1-n_1$ is even and $m_2-n_2$ is odd), we find that
$m_1+m_2=q$, and the two-point function in this case is hence analytic
in $\mu$ and is disregarded as nonuniversal.

If both differences $m_1-n_1$ and $m_2-n_2$ have the same parity,
the dimensional analysis shows that
\be
\langle O_{m_1 n_1} O_{m_2n_2}\rangle\sim \delta_{m_1,m_2}\mu^{\frac{m_1-n_1}{2}+\frac{m_2-n_2}{2}+\frac{m_1}{q}}.
\ee
Taking $m_1\in[1,q-1]$ into account, we see that the two-point function
cannot be analytic in $\mu$ and must therefore be equal to zero once the
pairs $(m_1,n_1)$ and $(m_2,n_2)$ are different. It proves useful to
introduce the new variable
\be\label{t}
t=2\bigg(\frac{v_1}{v_{1}^*}\bigg)^q-1.
\ee
In the case where both fields are even, the degree of the polynomial
$\frac{\partial S^{(m n )}}{\partial v_{m}}$ is
\be\label{partialS}
\frac{\partial S^{(m n )}}{\partial v_{m}}(t)\sim t^{\frac{m-n}{2}}+\dots,
\ee
where the dots denote terms subleading in $t$. The diagonality condition
for the two-point correlation function in terms of the new variable takes the form
\begin{equation}\label{even}
\langle O_{m_1 n_1} O_{m_2n_2}\rangle\sim \delta_{m_1,m_2} \int_{-1}^1 d t\,(1+t)^{\frac{m_1+m_2}{2q}-1}\,
\frac{\partial S^{(m_1n_1)}}{\partial v_{m_1}}\frac{\partial S^{(m_2n_2)}}{\partial v_{m_2}}=0.
\end{equation}
In the case where both fields are odd, it is convenient to change
$\frac{\partial S^{(m n)}}{\partial v_{\alpha}}=(1+t)^{\frac{m}{q}}
\frac{\partial \tilde S^{(m n)}}{\partial v_{\alpha}}$. The degree of the new polynomial is
\be \label{partialtildeS}
\frac{\partial \tilde S^{(m n)}}{\partial v_{\alpha}}\sim t^{\frac{m-n-1}{2}}+\dots,
\ee
and the diagonality condition is
\begin{equation}\label{odd}
\langle O_{m_1 n_1} O_{m_2n_2}\rangle\sim \delta_{m_1,m_2} \int_{-1}^1 d t\,(1+t)^{\frac{m_1+m_2}{2q}}\,
\frac{\partial \tilde S^{(m_1n_1)}}{\partial v_{q-m_1}}\frac{\partial \tilde S^{(m_2n_2)}}{\partial v_{q-m_2}}=0.
\end{equation}
Equations \eqref{even} and \eqref{odd} show that the diagonal form of the
two-point correlation numbers requires that the polynomials \eqref{partialS}
and \eqref{partialtildeS} form an orthogonal set of polynomials.
Such polynomials are known as Jacobi orthogonal polynomials.
In our case we are dealing with the special class of the Jacobi polynomials $P_n^{(0,b)}(t)$.
They are $n$th order polynomials that form an orthogonal system on the
interval $[0,1]$ with the orthogonality condition
\begin{equation}
\int_{-1}^{1} d t (1+t)^b P_n^{(0,b)}(t)P_m^{(0,b)}(t)=\frac{2^{b+1}}{2n+b+1}\delta_{m,n}.
\label{JacobiScalar}
\end{equation}
In the standard normalization, $P_n^{(0,b)}(1)=1$.
We note that the Jacobi polynomials in this normalization
have the highest coefficient
\begin{equation}
P_n^{(0,b)}(t) =\frac{(b+n+1)_{n}}{n!}\, \bigg(\frac{t}{2}\bigg)^n+\ldots.
\end{equation}
Summarizing these results, we conclude that
\begin{eqnarray}\label{Smn_alpha}
\frac{\partial S^{(mn)}}{\partial v_{\alpha}}(v_1)=\begin{cases}
\delta_{m,\alpha} v_{*1}^{\frac{m-n}{2}q} N_{mn} P_{\frac{m-n}{2}}^{(0,\frac{m-q}{q})}(t),&(m-n)\text{ even},\\
\delta_{m,q-\alpha} v_{*1}^{\frac{m-n-1}{2}q+m} N_{mn} (1+t)^{\frac{m}{q}}
P_{\frac{m-n-1}{2}}^{(0,\frac{m}{q})}(t),&(m-n)\text{ odd},
\end{cases}
\label{Smn}
\end{eqnarray}
where $t$ is defined in \eqref{t} and $N_{mn}$ is the normalization factor independent of $t$.
Although we obtained \eqref{Smn_alpha} for $v_{i>1}=0$, the general form of $S^{(mn)}$
can be easily reconstructed using equation \eqref{Smn0}. Hence, we have the answer for the 
action $S$ defined in \eqref{actionlambda} up to second order in the coupling constant expansion.
This result is required for calculating multipoint correlation functions.

\section{Conclusions}
\label{sec:Concl}

In this paper, we presented the solution of the Douglas string equation and the explicit form of the resonance relations up to second order in the coupling constant expansion, which lead to satisfying the main requirements of minimal Liouville gravity, namely, the requirements that the one-point correlation function except the unity operator must be zero and that the two-point correlators must be diagonal. Using the connection between the Douglas approach and the Frobenius manifold structure, we showed that the appropriate solution of the string equation has a remarkably simple form in the flat coordinates. Calculations of the multipoint correlation functions require this information. In addition, it requires knowing the structure constants of the Frobenius algebra. We found the structure constants in the canonical coordinates as well as the expression for the structure constants on the solution of the string equation in the flat coordinates, i.e., for all except the first coordinates be zero.  To obtain the general answer for the partition function, which is appropriate for calculating arbitrary multipoint correlators, the explicit form of the general structure constants in the flat coordinates is required. The general form of the resonance relations is also needed. Presumably, higher terms in the coupling constant decomposition are fixed from the restrictions following from the fusion rules. We 
are going to study this question in the near future.

\newpage
\noindent \textbf{Acknowledgements.} I am grateful to  A.~Belavin
for useful comments and discussions and K.~Alkalaev for comments on the draft of this article. 
I thank G.~Mussardo, SISSA and K.~Narain, ICTP, Trieste, Italy for the
hospitality during my visits in 2013. The research was performed 
under a grant funded  by Russian Science Foundation 
(project No. 14-12-01383).

\vspace{5mm}


\begin{appendices}
\section{$C_{\alpha\beta\gamma}$ }
If two of the indices are raised, then the structure constants in $\bf{u}$
coordinates is known from \eqref{strconst}. On the other hand,
the metric in flat coordinates is simple, and lowering an index $\alpha$
hence is just replacing it with $q-\alpha$. Therefore,
\be
C_{\alpha\beta\gamma} =\frac{\partial v^{q-\alpha}}{\partial u^{i}}
\frac{\partial v^{q-\beta}}{\partial u^{j}}\frac{\partial u^k}{\partial v^{\gamma}} \widetilde{C}_{k}^{ij}.
\ee
Using the explicit form of the coordinate transformation \eqref{transform},
we find that on the line $v_{k>1}=0$, we have
\begin{eqnarray}
\frac{\partial u^k}{\partial v^{\gamma}}({\bf{v}})= {U}_{\gamma}^k,\\
\frac{\partial v^\alpha}{\partial u^{i}}({\bf{v}})= {V}_{\gamma}^k,
\end{eqnarray}
where
\begin{align}
&{U}_{\gamma}^k=\binom{\frac{\gamma-k+q-2}{2}}{\frac{\gamma+k-q}{2}}\bigg(\frac{v_1}{2}\bigg)^{\frac{\gamma+k-q}{2}} &\text{ if }&\quad\frac{\gamma+k-q}{2}\in \mathbb{N},
\quad &\text{otherwise  }0, \label{U}
\\
&{V}^{\rho}_j=\frac{2\rho}{q+\rho-j}\binom{q-j-1}{\frac{q-\rho-j}{2}}
\bigg(\!\!-\frac{v_1}{2}\bigg)^{\frac{q-\rho-j}{2}} &\text{ if }&\quad\frac{q-\rho-j}{2}\in \mathbb{N},
\quad & \text{otherwise  }0.
\end{align}
In particular, from \eqref{U}, we find
\be
u^k=\frac{2q}{k+1}\binom{\frac{2q-k-1}{2}}{\frac{k-1}{2}}
\bigg(\frac{v_1}{2}\bigg)^{\frac{k+1}{2}} \text{ if }\quad\frac{k+1}{2}\in
\mathbb{N},
\quad \text{otherwise  }0.
\ee
In this notation, we find
\begin{align}
&C_{\alpha\beta\gamma}({\bf v}_*)={V}^{q-\alpha}_j{V}^{q-\beta}_j{U}_{\gamma}^k\widetilde{C}_{k}^{ij}.
\end{align}
Using standard identities for the binomial coefficients, we obtain \eqref{strconstflat}.

\newpage
\section{Evaluation of $\frac{\partial \theta_{\lambda,k}}{\partial v_{\alpha}}$}

We use our basic recurrence relation
\begin{equation}
\frac{\partial^2 \theta_{\lambda,k}}{\partial v_1 \partial v_{\alpha}}=
C_{\gamma}^{1\alpha} \frac{\partial \theta_{\lambda,k-1}}{\partial v_{\gamma}}.
\end{equation}
We integrate this equation over $v_1$ to find
\begin{equation}
\frac{\partial \theta_{\lambda,k}}{ \partial v_{\alpha}}=
\int^{v_1} C_{\gamma}^{1\alpha} \frac{\partial \theta_{\lambda,k-1}}{\partial v_{\gamma}} d v_1,
\end{equation}
where the integration constant is absent because it is independent of $v_1$
and can hence depend on only $v_{\alpha>1}$, which are set to zero.
Using this equation iteratively, we find
\begin{equation}
\frac{\partial \theta_{\lambda,k}}{ \partial v_{\alpha}}=
\int^{v_1}\!\!\!d v_1^{(1)}\dots\int^{v_1^{(k-1)}} \!\!\!\!\!\!\!\!\!d v_1^{(k)}\,\,
C_{\gamma_1}^{1\alpha}(v_1^{(1)})C_{\gamma_2}^{1\gamma_1}(v_1^{(2)})\dots C_{\gamma_k}^{1\gamma_{k-1}}(v_1^{(k)})
\,\, \frac{\partial \theta_{\lambda,0}}{\partial v_{\gamma_k}},
\end{equation}
where explicitly $\frac{\partial \theta_{\lambda,0}}{\partial v_{\gamma_k}}=\delta_{\lambda,\gamma_k}$.
From the explicit form of the structure constant \eqref{strconstflat}, we conclude that
\begin{eqnarray}
&&\gamma_1=\gamma_3=\dots=q-\alpha,\\
&&\gamma_2=\gamma_4=\dots=\alpha.
\end{eqnarray}
In particular, we find that $\lambda=\alpha$ for even $k$ while
$\lambda=q-\alpha$ for odd $k$. Otherwise, the integral is equal to zero.
After contracting all the indices, we are left with a multiple integral
containing
\be
C_{q-\alpha}^{1\alpha}(v_1^{(i)})=\bigg(\!\!-\frac{v_1^{(i)}}{q}\bigg)^{q-\alpha-1}
\ee
and
\be
C_{\alpha}^{1\, q-\alpha}(v_1^{(i)})=\bigg(\!\!-\frac{v_1^{(i)}}{q}\bigg)^{\alpha-1}.
\ee
Explicit integration gives
\begin{eqnarray}
\begin{cases}
k\text{ even}:&
\frac{\partial \theta_{\lambda,k}}{\partial v_{\alpha}}=
\delta_{\lambda,\alpha}
\big(\!\!-\frac{1}{q}\big)^{\frac{k}{2}(q-2)}
\frac{v_1^{\frac{k}{2}q}}{\alpha q (\alpha+q)2q(\alpha+2q)3q\dots
(\alpha+(\frac{k}{2}-1)q)(\frac{k}{2}q)},
\\
k\text{ odd}:&
\frac{\partial \theta_{\lambda,k}}{\partial v_{\alpha}}=
\delta_{\lambda,q-\alpha}
\big(\!\!-\frac{1}{q}\big)^{\frac{k-1}{2}(q-2)+q-\alpha-1}
\!\!\frac{v_1^{\frac{k+1}{2}q-\alpha}}{(q-\alpha)q(2q-\alpha)2q(3q-\alpha)3q\dots
(\frac{k-1}{2}q)(\frac{k+1}{2}q-\alpha)}.
\end{cases}
\end{eqnarray}
This gives \eqref{Lemma3} and \eqref{Lemma31}.

\end{appendices}

\newpage

\providecommand{\href}[2]{#2}\begingroup\raggedright
\addtolength{\baselineskip}{-3pt} \addtolength{\parskip}{-1pt}

\end{document}